# Band offsets and electronic structure at α/β-$Ga_2O_3$ and κ/β-$Ga_2O_3$ phase heterojunctions

D. Q. Fang*

MOE Key Laboratory for Nonequilibrium Synthesis and Modulation of Condensed Matter, School of Physics, Xi'an Jiaotong University, Xi'an 710049, China

Abstract

$Ga_2O_3$ phase heterojunction materials are emerging as promising candidates for deep-ultraviolet optoelectronics. Here, the band offsets at the α/β-$Ga_2O_3$ and κ/β-$Ga_2O_3$ phase heterojunctions are investigated through hybrid functional calculations. Explicit, realistic slab models that satisfy electron counting rules are constructed for the heterojunctions. Our calculations demonstrate that a type-II staggered band alignment forms at the α/β-$Ga_2O_3$ and κ/β-$Ga_2O_3$ phase heterojunctions, where both the valence and conduction band edges of α-$Ga_2O_3$ and κ-$Ga_2O_3$ are higher than those of β-$Ga_2O_3$. Strain effects on the band alignment are also discussed. Spatially resolved local density of states maps are presented, clearly revealing the interfacial electronic structures. These results provide valuable insights into the physical properties of $Ga_2O_3$ phase heterojunctions, offering guidance for future device design efforts.

*fangdqphy@xjtu.edu.cn

## I. Introduction

Gallium oxide ($Ga_2O_3$) is an ultrawide-bandgap semiconductor, which adopts a monoclinic (β) phase in its ground state. The fundamental bandgap of β-$Ga_2O_3$ is about 4.85 eV [1]. The ultrawide bandgap energy of β-$Ga_2O_3$ allows its applications in deep-ultraviolet (UV) optoelectronics such as solar-blind photodetectors. Its large critical electric field strength and the ability to achieve controllable n-type doping make it suitable for high power switching applications such as field effect transistors [2,3] and Schottky Barrier diodes [4].

Although β-$Ga_2O_3$ holds great promise for next-generation devices, the lack of p-type conductivity remains a major drawback for its optoelectronic applications [5]. Constructing heterostructures offers a viable strategy to circumvent this limitation. For instance, materials such as CuI [6], $TiO_2$ [7], ZnO [8], $CuGaO_2$ [9], and GaN [10] have been integrated with β-$Ga_2O_3$ to fabricate UV photodetectors. However, because these secondary materials possess narrower bandgaps than β-$Ga_2O_3$, the resulting devices typically exhibit a broad-spectrum UV response, rendering them suboptimal for solar-blind photodetection.

$Ga_2O_3$ exhibits multiple polymorphs (i.e., α, β, γ, κ, and δ) with comparable bandgaps [11] and readily undergoes phase transitions [12]. Recently, $Ga_2O_3$ phase heterojunctions have attracted great attention for use in solar-blind photodetectors [13–16]. Guo et al. reported the fabrication of α/β-$Ga_2O_3$ phase junction nanorod arrays (NRAs) with a thickness-controllable β-$Ga_2O_3$ shell layer via hydrothermal and postannealing treatments, and demonstrated that photodetectors based on these NRAs show self-powered and solar-blind photoelectric characteristics [13]. Hu et al fabricated a solar-blind photodetector based on α/β-$Ga_2O_3$ phase junction NRAs using $Ti_3C_2$-siliver nanowires as the top electrode [15]. Under 254 nm light at a power density of 100 μW $cm^{-2}$ and zero bias, the device exhibited a large detectivity of $1.45 \times 10^{14}$ Jones and a responsivity of 5.2 mA $W^{-1}$. κ/β-$Ga_2O_3$ phase heterojunction film was also demonstrated experimentally [16]. The photodetector fabricated using the κ/β-$Ga_2O_3$ phase heterojunction film displayed a detectivity of $1.69 \times 10^{10}$ Jones and a responsivity of 17.8 mA $W^{-1}$ under 240 nm light at a power density of 11.3 μW $cm^{-2}$ and zero bias.

Band alignment between different materials or phases is crucial for heterostructure design and for understanding the transfer mechanisms of photogenerated electrons and holes [17,18]. In this work, we construct the atomic structures of α/β-$Ga_2O_3$ and κ/β-$Ga_2O_3$ phase heterojunctions and investigate their band offsets and electronic structures using first-principles calculations. The paper is organized as follows. Section II introduces the computational details of the electronic structure and band offsets, and describes the construction of interface models. In Section III, the interface properties, band alignment, and local density of states are presented and analyzed, together with a comparison with available experimental and theoretical results. In Section IV, we summarize our results.

## II. Computational details

### A. First-principles calculations

Our first-principles calculations are carried out within the framework of density functional theory, using the projector augmented-wave method [19,20], as implemented in the Vienna *ab initio* Simulation Package [21]. Exchange-correlation effects are described by a revised Perdew-Burke-Ernzerhof generalized gradient approximation for solids (PBEsol) [22]. The Ga 3*d* electrons are treated explicitly as valence electrons. To obtain strain-free bulk structures, both atomic coordinates and cell parameters are optimized using a cutoff energy of 520 eV for the plane-wave basis set. Brillouin zone (BZ) sampling is performed through 6×6×2, 3×9×6, and 5×3×3 Γ-centered **k**-point meshes for the unit cells of α-, β-, and κ-$Ga_2O_3$, including 30 atoms, 20 atoms, and 40 atoms, respectively. For the structural relaxations of the interface models, a 520 eV plane-wave cutoff energy and a 2×3×1 Γ-centered **k**-point mesh are used. To evaluate integrals over the Brillouin zone, the Gaussian smearing is applied with a width of 0.05 eV. All atoms are relaxed until the Hellmann-Feynman forces are less than 0.005 eV $Å^{-1}$ and 0.02 eV $Å^{-1}$ for the bulk and interface cases, respectively.

To address the bandgap underestimation problem of the PBEsol functional, standard PBE0 hybrid functional calculations [23] are performed at PBEsol-relaxed geometries. To reduce the computational load, a plane-wave cutoff energy of 400 eV is adopted for all PBE0 calculations, which maintains the accuracy of the $Ga_2O_3$ bulk energy level within 0.01 eV relative to a 520 eV cutoff. A Γ-centered **k**-point mesh of 1×2×1 is employed for the interface PBE0 calculations due to the large slab size (≥ 570 atoms).

B. Interface models

Experimentally, the growth of β-$Ga_2O_3$ on α-$Ga_2O_3$ was found to exhibit an epitaxial relationship of (0001) α-$Ga_2O_3$ ‖ $(\bar{2}01)$ β-$Ga_2O_3$ [24,25]. Lu et al. [16] successfully fabricated a κ-$Ga_2O_3$/β-$Ga_2O_3$ phase heterojunction film featuring a sharp and well-defined interface, which exhibits an epitaxial relationship of (001) κ-$Ga_2O_3$ ‖ $(\bar{2}01)$ β-$Ga_2O_3$.

To study the atomic and electronic structures of the α/β-$Ga_2O_3$ and κ/β-$Ga_2O_3$ phase interfaces, we model the systems as finite slabs separated by a vacuum region larger than 12 Å in supercells. For the α/β-$Ga_2O_3$ phase heterojunction, an out-of-plane epitaxial relationship of (0001) α-$Ga_2O_3$ ‖ $(\bar{2}01)$ β-$Ga_2O_3$ is employed, in agreement with experimental observations, and the in-plane epitaxial relationships are $[2\bar{1}\bar{1}0]$ α-$Ga_2O_3$ ‖ [102] β-$Ga_2O_3$ and $[01\bar{1}0]$ α-$Ga_2O_3$ ‖ [010] β-$Ga_2O_3$. We construct an orthorhombic supercell of 570 atoms for the α/β-$Ga_2O_3$ heterostructure comprising a α-$Ga_2O_3$ slab with (3×1) in-plane periodicity and a β-$Ga_2O_3$ slab with (1×3) in-plane periodicity. For the κ/β-$Ga_2O_3$ phase heterojunction, we employ the out-of-plane epitaxial relationship of (001) κ-$Ga_2O_3$ ∥ ($\bar{2}01$) β-$Ga_2O_3$ and the in-plane epitaxial relationships of [100] κ-$Ga_2O_3$ ∥ [102] β-$Ga_2O_3$ and [010] κ-$Ga_2O_3$ ∥ [010] β-$Ga_2O_3$. An orthorhombic supercell containing 600 atoms is constructed for the κ/β-$Ga_2O_3$ heterostructure, with the in-plane periodicities of (3×1) for the κ-$Ga_2O_3$ slab and (1×3) for the β-$Ga_2O_3$ slab. For the interface models, the $x$ and $y$ axes are parallel to the [102] and [010] crystal axes of β-$Ga_2O_3$, respectively, and the $z$ axis is normal to the ($\bar{2}01$) plane of β-$Ga_2O_3$.

C. Band alignment calculations

The band alignment is evaluated using a coherent interface model alongside two corresponding bulk calculations. The coherent interface model inherently incorporates strain into at least one of the constituent materials. The reference level offset is derived directly from the interface calculation. In the bulk calculations, to account for strain effects on the valence band maximum (VBM) and conduction band minimum (CBM) energy positions, the in-plane lattice parameters (ILPs) are constrained to match those of the interface model, while the remaining lattice parameters and internal atomic coordinates are optimized.

The macroscopically averaged electrostatic potential is chosen as the reference level, which is defined as [26]

$$\bar{\bar{V}}(z) = \frac{1}{a}\int_{z-a/2}^{z+a/2} \bar{V}(z')dz' , \quad (1)$$

where $\bar{V}(z)$ and $a$ are the planar-averaged electrostatic potential and the period length of material, respectively. To concurrently extract the macroscopically averaged electrostatic potentials for both phase A and phase B within the A/B heterojunction model, we employ a modified version of the MacroDensity code [27], with the period length explicitly set to the respective value of each phase.

The valence band offset (VBO) is calculated as [28]

$$\Delta\varepsilon_{VBM,X}^{A-B} = \Delta\varepsilon_{VBM-\mathrm{Re}f,X}^{A} + \Delta\varepsilon_{\mathrm{Re}f,X}^{A-B} - \Delta\varepsilon_{VBM-\mathrm{Re}f,X}^{B}, \quad (2)$$

where $\Delta\varepsilon_{\mathrm{Re}f,X}^{A-B}$ denotes the reference level offset derived from the interface model with the ILPs of set X, and $\Delta\varepsilon_{VBM-\mathrm{Re}f,X}^{A}$ ($\Delta\varepsilon_{VBM-\mathrm{Re}f,X}^{B}$) is the VBM position relative to the reference level for phase A (B) at the ILPs of set X, determined through bulk calculation. The conduction band offset (CBO) is obtained by adding the difference of the calculated bandgaps of phases A and B to the VBO.

## III. Results and discussions

### A. Bulk properties of $Ga_2O_3$ polymorphs

Figure 1 depicts the unit cells of α-, κ-, and β-$Ga_2O_3$ employed for phase heterojunction construction. α-$Ga_2O_3$ crystallizes in a corundum structure (space group $R\bar{3}c$) [Fig. 1(a)], with Ga atoms only in a sixfold coordination environment. The orthorhombic κ-$Ga_2O_3$ polymorph with space group $Pna2_1$ exhibits a complex structure wherein Ga atoms possess fourfold, fivefold, and sixfold coordination configurations [11], as shown in Fig. 1(b). The monoclinic β-$Ga_2O_3$ polymorph with space group $C2/m$ features Ga atoms in fourfold and sixfold coordination configurations [Fig. 1 (c)]. Table I and Table II present the calculated lattice parameters and bandgaps, respectively, for α-, β-, and κ-$Ga_2O_3$. The lattice parameters predicted by PBEsol show good agreement with experimental data, whereas the bandgaps are significantly underestimated. Calculations using the PBE0 hybrid functional yield theoretical bandgaps that closely approach experimental values.

B. Interface properties of $Ga_2O_3$ phase heterojunctions

To study the strain effects on the band offsets of $Ga_2O_3$ phase heterojunctions, we choose different $Ga_2O_3$ phases as the substrate. The lattice strain is defined as $\varepsilon = (a - a_0)/a_0$, where $a_0$ and $a$ are the unstrained and strained lattice parameters, respectively. The in-plane lattice strains of the individual phases in the A/B heterojunction are listed in Table III. In the A/B heterojunction, A represents the substrate and B denotes the deposited film. For the α-$Ga_2O_3$/β-$Ga_2O_3$ heterojunction, the β-$Ga_2O_3$ film experiences a tensile strain of 1.68% along the $x$-direction and a compressive strain of −5.26% along the $y$-direction. Conversely, in the case of the β-$Ga_2O_3$/α-$Ga_2O_3$ heterojunction, the α-$Ga_2O_3$ film is subjected to a compressive strain of −1.65% along the $x$-direction and a tensile strain of 5.55% along the $y$-direction. A similar strain behavior is also observed in the junctions formed by κ-$Ga_2O_3$ and β-$Ga_2O_3$.

Figures 2(a) and 2(b) show the relaxed atomic structures of the (0001) α-$Ga_2O_3$ ∥ $(\bar{2}01)$ β-$Ga_2O_3$ and (001) κ-$Ga_2O_3$ ∥ $(\bar{2}01)$ β-$Ga_2O_3$ heterojunctions, respectively, where the interfacial Ga atoms possess a tetrahedrally coordinated environment. Importantly, the proposed interface models adhere to the electron counting rule [29]. Taking the α-$Ga_2O_3$/β-$Ga_2O_3$ interface model as an example, Ga and O layers alternate along the direction perpendicular to the interface. Around the interface, each O layer consists of 18 $O^{2-}$ ions (−36 charges), which are exactly neutralized by the adjacent Ga layer containing 12 $Ga^{3+}$ ions (+36 charges).

C. Band alignment of $Ga_2O_3$ phase heterojunctions

Figure 3(a) plots the planar-averaged and macroscopically averaged electrostatic potentials calculated at the PBE0 level for the α-$Ga_2O_3$/β-$Ga_2O_3$ heterojunction. The macroscopically averaged electrostatic potentials in the bulk-like regions of α-$Ga_2O_3$ and $\beta$-$Ga_2O_3$ are flat, and thus the reference level offset is directly determined to be −0.20 eV, as listed in Table IV, where the negative sign indicates that the reference level of $\alpha$-$Ga_2O_3$ lies below that of $\beta$-$Ga_2O_3$. The band edge positions of $\alpha$-$Ga_2O_3$ relative to the reference level are obtained from the PBE0 calculation on an unstrained bulk unit cell. To determine the band edge positions of strained $\beta$-$Ga_2O_3$, a larger transformed unit cell of $\beta$-$Ga_2O_3$ with the $(\bar{2}01)$ facet is employed [Fig. 1(d)], wherein its first lattice vector is $\boldsymbol{a}' = \boldsymbol{a} + 2\boldsymbol{c}$, with $\boldsymbol{a}$ and $\boldsymbol{c}$ representing the lattice vectors of the conventional unit cell of $\beta$-$Ga_2O_3$, while the other two lattice vectors remain unchanged and the angle between $\boldsymbol{a}'$ and $\boldsymbol{c}$ is defined as $\beta'$ [30]. To account for lattice matching with the $\alpha$-$Ga_2O_3$ substrate, the lattice parameters $a'$ and $b$ of this transformed unit cell are strained accordingly, and the other lattice parameters ($c$ and $\beta'$) and atomic coordinates are optimized. Electronic structure calculations using the PBE0 functional yield the band edge positions of strained $\beta$-$Ga_2O_3$ relative to the reference level. According to Equation (2), the VBO and CBO for the $\alpha$-$Ga_2O_3$/$\beta$-$Ga_2O_3$ heterojunction are calculated to be 0.31 eV and 0.56 eV, indicating that a type-II band alignment is formed between the $\alpha$-$Ga_2O_3$ substrate and the $\beta$-$Ga_2O_3$ film [Fig. 4(a)]. Consequently, photogenerated electrons transfer from $\alpha$-$Ga_2O_3$ to $\beta$-$Ga_2O_3$, while holes migrate from $\beta$-$Ga_2O_3$ to $\alpha$-$Ga_2O_3$.

As shown in Fig. 4(b), for the $\beta$-$Ga_2O_3$/$\alpha$-$Ga_2O_3$ heterojunction grown on the $\beta$-$Ga_2O_3$ substrate, the conduction and valence band edges of $\alpha$-$Ga_2O_3$ also lie higher in energy than those of $\beta$-$Ga_2O_3$. While the magnitude of the VBO remains unchanged, that of the CBO is reduced by 0.32 eV.

$\kappa$-$Ga_2O_3$ is a ferroelectric wide-bandgap semiconductor [31]. Within the (001) $\kappa$-$Ga_2O_3$ ∥ $(\bar{2}01)$ $\beta$-$Ga_2O_3$ interface model, the polar nature of the (001) $\kappa$-$Ga_2O_3$ slab gives rise to finite electric field across the interface under the periodic boundary condition. Consequently, the macroscopically averaged electrostatic potentials exhibit a distinct tilt, as show in Fig. 3(b). To obtain the reference level offset not influenced by the electric fields, the macroscopically averaged electrostatic potential for each phase in the slab is extrapolated from the bulk-like region to the nominal interface position. The midpoint between the surface Ga layer of κ-$Ga_2O_3$ and that of β-$Ga_2O_3$ is adopted as the nominal interface position, as shown in Fig. 2(b). This extrapolation scheme has been successfully applied to the interfaces between β-$Ga_2O_3$ and wurtzite III-N (III = Al, Ga) [30]. The reference level offset between κ-$Ga_2O_3$ and β-$Ga_2O_3$ is evaluated to be 0.48 eV, as listed in Table IV. By incorporating the band-edge energy positions of κ-$Ga_2O_3$ and strained β-$Ga_2O_3$, the VBO and CBO for the κ-$Ga_2O_3$/β-$Ga_2O_3$ heterojunction are calculated to be 0.67 eV and 0.66 eV, respectively, indicating a type-II band alignment [Fig. 4(c)]. For the β-$Ga_2O_3$/κ-$Ga_2O_3$ heterojunction grown on the β-$Ga_2O_3$ substrate, the magnitudes of the VBO and CBO decrease by 0.14 eV and 0.27 eV, respectively, as shown in Fig. 4(d).

D. Local density of states

To gain more insight into the electronic structures of the $Ga_2O_3$ phase interfaces, we calculate the in-plane integrated local density of states (LDOS) $\overline{L}(E,z)$ using the DensityTool code [32,33]:

$$\overline{L}(E,z)=\frac{\Omega_{cell}}{(2\pi)^3}\sum_n\int_{BZ}\delta(E-\varepsilon_{n,\mathbf{k}})\overline{P}_{n,\mathbf{k}}(z)d^3k\,, \quad (3)$$

where $\Omega_{cell}$ is the unit cell volume, $E$ is the energy, and $\overline{P}_{n,\mathbf{k}}(z)$ is the in-plane integrated partial charge density $P_{n,\mathbf{k}}(\mathbf{r})=|\varphi_{n,\mathbf{k}}(\mathbf{r})|^2$ with the Kohn-Sham wave functions $\varphi_{n,\mathbf{k}}(\mathbf{r})$ and eigenvalues $\varepsilon_{n,\mathbf{k}}$.

Owing to the large supercell size, LDOS calculations are performed using PBEsol.

Figure 5(a) displays the LDOS for the α-$Ga_2O_3$/β-$Ga_2O_3$ heterojunction. In the bulk-like regions, both the valence band edge and the conduction band edge of α-$Ga_2O_3$ are higher in energy than those of β-$Ga_2O_3$, qualitatively consistent with Fig. 4(a), even though the semi-local PBEsol functional is used here. Moreover, the interface states that emerge at the top of the valence band are distributed around the interface. It is noted that LDOS is usually not suitable for a precise extraction of the band offset since (i) states originating from one phase or the interface often extend into the band gap of the second phase, making the recognition of band edges complicated, and (ii) the δ function in Eq. (3) is substituted by a Gaussian with a smearing parameter, which can introduce errors in the determination of energy levels [32,33].

The LDOS for the κ-$Ga_2O_3$/β-$Ga_2O_3$ heterojunction is shown in Fig. 5(b). The presence of the electric field noticeably tilts the band edges of κ-$Ga_2O_3$. Furthermore, the lowest conduction band on one side of the slab falls below the Fermi level and is thus partially occupied, which is accompanied by unoccupied valence band states near the interface. Band tilting prevents the direct identification of the band alignment type from the LDOS. Therefore, eliminating band tilting is essential to qualitatively assess the band alignment of the κ-$Ga_2O_3$/β-$Ga_2O_3$ heterojunction.

Inspired by the scheme proposed by Yoo et al. [34], in which pseudo-hydrogen (psH) atoms were employed to passivate materials with spontaneous polarization to eliminate the macroscopic field, we terminate the surface Ga atoms on one side of the slab using psH atoms of valence 1.66, as shown in Fig. 5(c). The psH-Ga bonding orbital lies below the VBM and is fully occupied by two electrons, leading to the creation of hole states within the valence band. For the psH passivated κ-$Ga_2O_3$/β-$Ga_2O_3$ heterojunction, the Fermi level enters the valence band and the band tilting disappears. From the LDOS in Fig. 5(c), it is observed that both the valence band edge and the conduction band edge of κ-$Ga_2O_3$ lie above those of β-$Ga_2O_3$, demonstrating the formation of the type-II band alignment.

E. Comparison with available experimental and theoretical results

The calculated band offsets together with available experimental and theoretical results are summarized in Table IV. Benz et al. study the band alignment of all four combinations of $Cu_2O$ and CuO with α-$Ga_2O_3$ and β-$Ga_2O_3$ using x-ray photoelectron spectroscopy and derive the VBO of 0.43±0.05 eV for α-$Ga_2O_3$ to β-$Ga_2O_3$ [35]. Our calculated VBO of 0.31 eV for the α-$Ga_2O_3$/β-$Ga_2O_3$ heterojunction is close to the experimental data. Ju et al. [36] constructed (110) α-$Ga_2O_3$ ∥ $(3\bar{1}0)$ β-$Ga_2O_3$ heterojunctions and found that the VBM and CBM of α-$Ga_2O_3$ lie 0.35 eV and 0.07 eV, respectively, above those of β-$Ga_2O_3$ based on the Heyd-Scuseria-Ernzerhof (HSE) hybrid functional calculations. Our VBO result is consistent with that reported by Ju et al. [36], while the discrepancy in CBO could be attributed to the epitaxial relationships, strain, and chemical bonding at the interface. Peelaers et al. [37] calculated a VBO of −0.03 eV and a CBO of 0.44 eV between unstrained α-$Ga_2O_3$ and unstrained β-$Ga_2O_3$ by aligning the band edges to the vacuum level. The discrepancies with our results originate from our explicit consideration of strain effects and interfacial chemical bonding. For the κ-$Ga_2O_3$/β-$Ga_2O_3$ heterojunction, Lu et al. experimentally reported a VBO of 0.65 eV and a CBO of 0.71 eV [16]. Our calculated results are in good agreement with the experimental data.

## IV. Conclusions

In summary, we investigate the band offsets at the α/β-$Ga_2O_3$ and κ/β-$Ga_2O_3$ phase heterojunctions through PBE0 hybrid functional calculations. All phase heterojunctions studied exhibit a type-II band alignment that facilitates the separation of photogenerated electrons and holes. For the α-$Ga_2O_3$/β-$Ga_2O_3$ phase heterojunction grown on the α-$Ga_2O_3$ substrate, the VBO and CBO are calculated to be 0.31 eV and 0.56 eV, respectively, with both the valence and conduction band edges of α-$Ga_2O_3$ higher than those of β-$Ga_2O_3$. When choosing β-$Ga_2O_3$ as the substrate in the β-$Ga_2O_3$/α-$Ga_2O_3$ phase heterojunction, the magnitude of VBO remains unchanged, but that of CBO is reduced by 0.32 eV. The LDOS map clearly reveal the interfacial electronic structure, which is consistent with band offset calculations.

Due to the presence of finite electric fields in the κ-$Ga_2O_3$/β-$Ga_2O_3$ phase heterojunction, the reference level offset is determined by extrapolating the macroscopically averaged electrostatic potential from the bulk-like regions to the nominal interface position. Using this scheme, the VBO and CBO for the κ-$Ga_2O_3$/β-$Ga_2O_3$ phase heterojunction are calculated to be 0.67 eV and 0.66 eV, respectively, which are in good agreement with the experimental results. For the β-$Ga_2O_3$/κ-$Ga_2O_3$ phase heterojunction grown on the β-$Ga_2O_3$ substrate, the magnitudes of the VBO and CBO reduce by 0.14 eV and 0.27 eV, respectively. The calculated band alignments are crucial for designing $Ga_2O_3$ phase-heterojunction optoelectronic devices.

## ACKNOWLEDGMENTS

The authors acknowledge the financial support from the National Natural Science Foundation of China (Grant Nos. 11604254 and 92265103) and the Innovation Program for Quantum Science and Technology (Grant No. 2021ZD0302400). The authors also acknowledge the HPCC Platform of Xi’an Jiaotong University for providing the computing facilities.

## Conflict of interest

The authors declare no conflict of interest.

## Data Availability Statement

The data that support the findings of this study are available from the corresponding author upon reasonable request.

Table I. Calculated and experimental lattice parameters (in Å) of α-, β-, and κ-$Ga_2O_3$.

| | α-$Ga_2O_3$ | | β-$Ga_2O_3$ | | κ-$Ga_2O_3$ | |
|---|---|---|---|---|---|---|
| | Calc. | Expt. [38] | Calc. | Expt. [39] | Calc. | Expt. [40] |
| *a* | 5.006 | 4.98 | 12.279 | 12.233 | 5.066 | 5.046 |
| *b* | | | 3.050 | 3.038 | 8.693 | 8.702 |
| *c* | 13.457 | 13.43 | 5.815 | 5.807 | 9.298 | 9.283 |
| β (°) | | | 103.75 | 103.82 | | |

Table II. PBE0 bandgaps (in eV) of α-, β-, and κ-$Ga_2O_3$ compared with the experimental values. The data calculated using PBEsol are given in parentheses.

| | Direct | Indirect | Experimental |
|---|---|---|---|
| α-$Ga_2O_3$ | 5.46 (2.92) | 5.22 (2.67) | 5.4 [25] |
| β-$Ga_2O_3$ | 4.84 (2.23) | 4.83 (2.21) | 4.85 [1] |
| κ-$Ga_2O_3$ | 4.90 (2.32) | 4.90 (2.32) | 4.9 [41] |

Table III. Strains in the in-plane lattice parameters $a_x$ and $a_y$ of the individual phases of the A/B heterojunction (A: substrate; B: film). A negative value indicates compressive strain, whereas a positive value indicates tensile strain.

| | α-$Ga_2O_3$ | | β-$Ga_2O_3$ | |
|---|---|---|---|---|
| Interface | $a_x$ | $a_y$ | $a_x$ | $a_y$ |
| α-$Ga_2O_3$/β-$Ga_2O_3$ | 0% | 0% | 1.68% | −5.26% |
| β-$Ga_2O_3$/α-$Ga_2O_3$ | −1.65% | 5.55% | 0% | 0% |

| | κ-$Ga_2O_3$ | | β-$Ga_2O_3$ | |
|---|---|---|---|---|
| Interface | $a_x$ | $a_y$ | $a_x$ | $a_y$ |
| κ-$Ga_2O_3$/β-$Ga_2$O3 | 0% | 0% | 2.90% | −5.01% |
| β-$Ga_2O_3$/κ-$Ga_2$O3 | −2.82% | 5.28% | 0% | 0% |

Table IV. Reference level difference $\Delta\varepsilon_{\mathrm{Re}f,X}^{A-B}$ (in eV) and band offsets (in eV) of the A/B heterojunction (A: substrate; B: film) calculated using the PBE0 functional. Available experimental and theoretical results are also provided. A positive value indicates that the reference level (or band edge) of substrate A lies above that of film B, while a negative value indicates that it lies below.

| Interface | $\Delta\varepsilon_{\mathrm{Re}f,X}^{A-B}$ | VBO | CBO |
|---|---|---|---|
| **α-$Ga_2O_3$/β-$Ga_2O_3$** | −0.20 | 0.31 | 0.56 |
| Expt. [35] | | 0.43±0.05 | |
| Calc. [36] | | 0.35 | 0.07 |
| Calc. [37] | | −0.03 | 0.44 |
| **β-$Ga_2O_3$/α-$Ga_2O_3$** | 0.29 | −0.31 | −0.24 |
| **κ-$Ga_2O_3$/β-$Ga_2O_3$** | 0.48 | 0.67 | 0.66 |
| Expt. [16] | | 0.65 | 0.71 |
| **β-$Ga_2O_3$/κ-$Ga_2O_3$** | −0.23 | −0.53 | −0.39 |

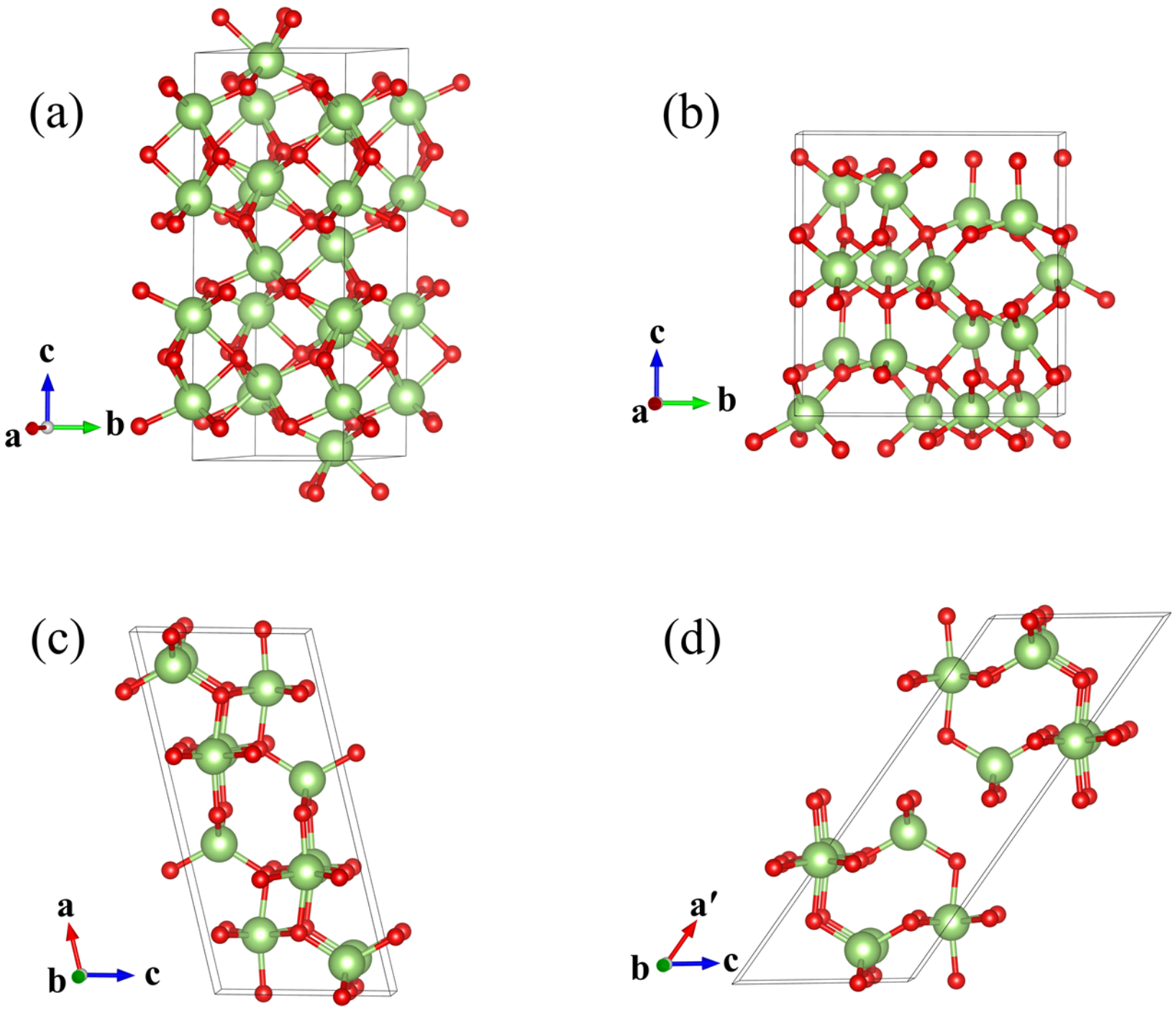

Figure 1. Top panel: unit cells of α-$Ga_2O_3$ (a) and κ-$Ga_2O_3$ (b). Bottom panel: conventional (c) and transformed (d) unit cells of β-$Ga_2O_3$. The relationship between the lattice vector $\boldsymbol{a}'$ of the transformed unit cell and the lattice vectors of the conventional unit cell is $\boldsymbol{a}' = \boldsymbol{a} + 2\boldsymbol{c}$.

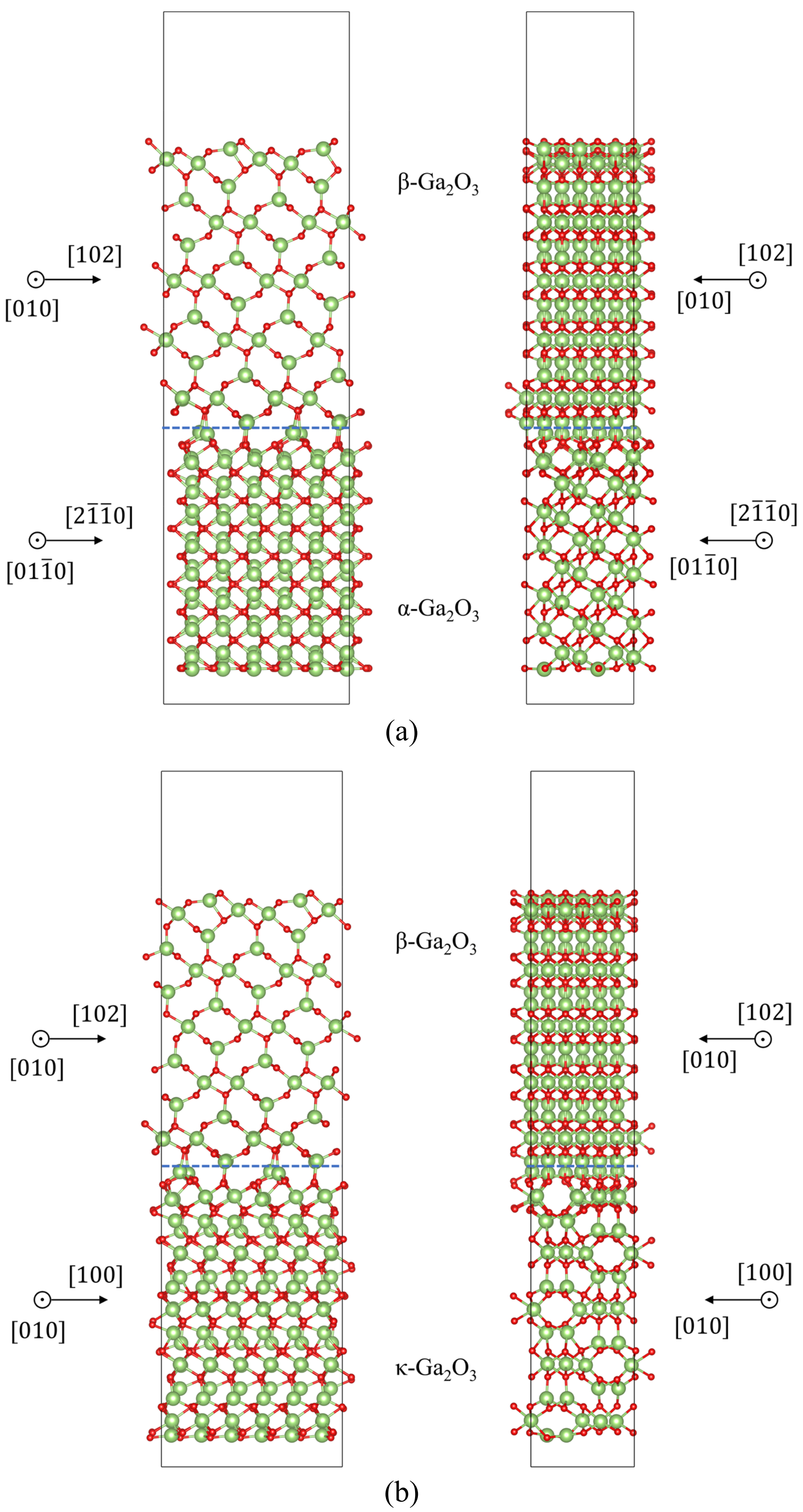

Figure 2. Relaxed atomic structures of the (0001) α-$Ga_2O_3$ ∥ ($\bar{2}$01) β-$Ga_2O_3$ (a) and (001) κ-$Ga_2O_3$ ∥ ($\bar{2}$01) β-$Ga_2O_3$ (b) heterojunctions. Blue dashed lines denote the interface positions.

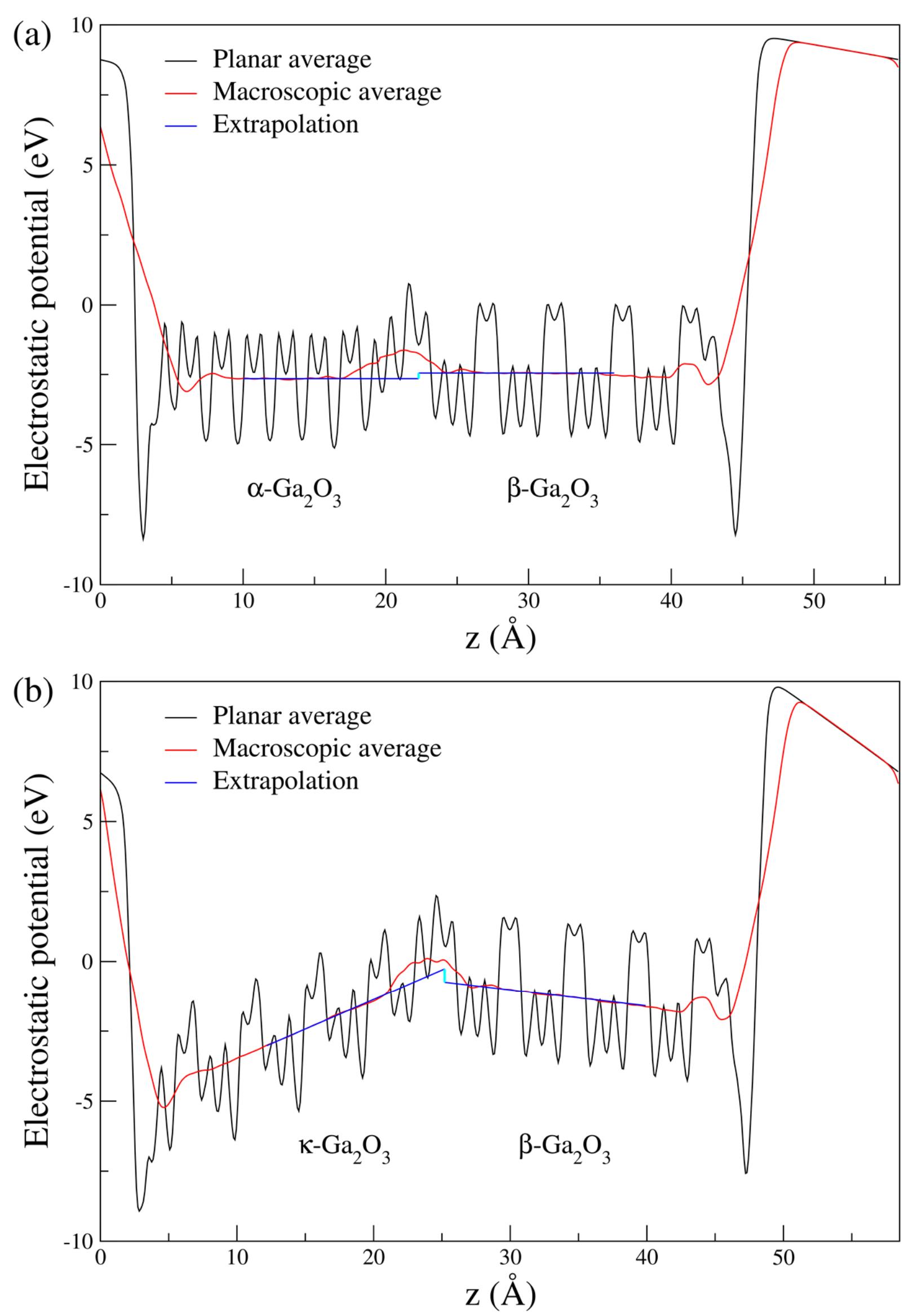

Figure 3. Planar-averaged and macroscopically averaged electrostatic potentials calculated at the PBE0 level for the α-$Ga_2O_3$/β-$Ga_2O_3$ (a) and κ-$Ga_2O_3$/β-$Ga_2O_3$ (b) heterojunctions.

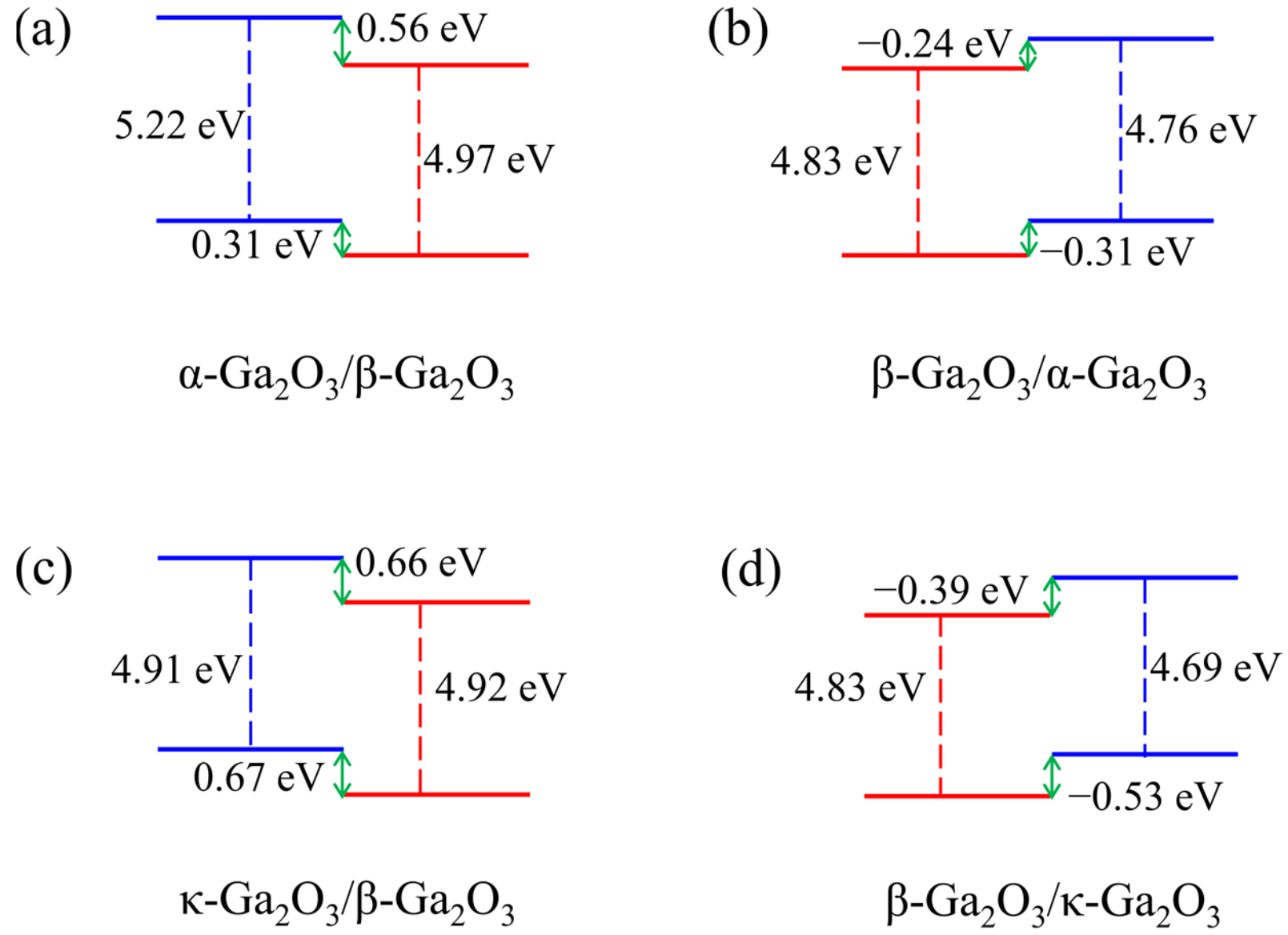

Figure 4. Band alignment of the interfaces between α-$Ga_2O_3$ and β-$Ga_2O_3$ and between κ-$Ga_2O_3$ and β-$Ga_2O_3$ calculated using the PBE0 hybrid functional. In (a) and (c), α-$Ga_2O_3$ and κ-$Ga_2O_3$ are the substrates, respectively. In (b) and (d), β-$Ga_2O_3$ is the substrate.

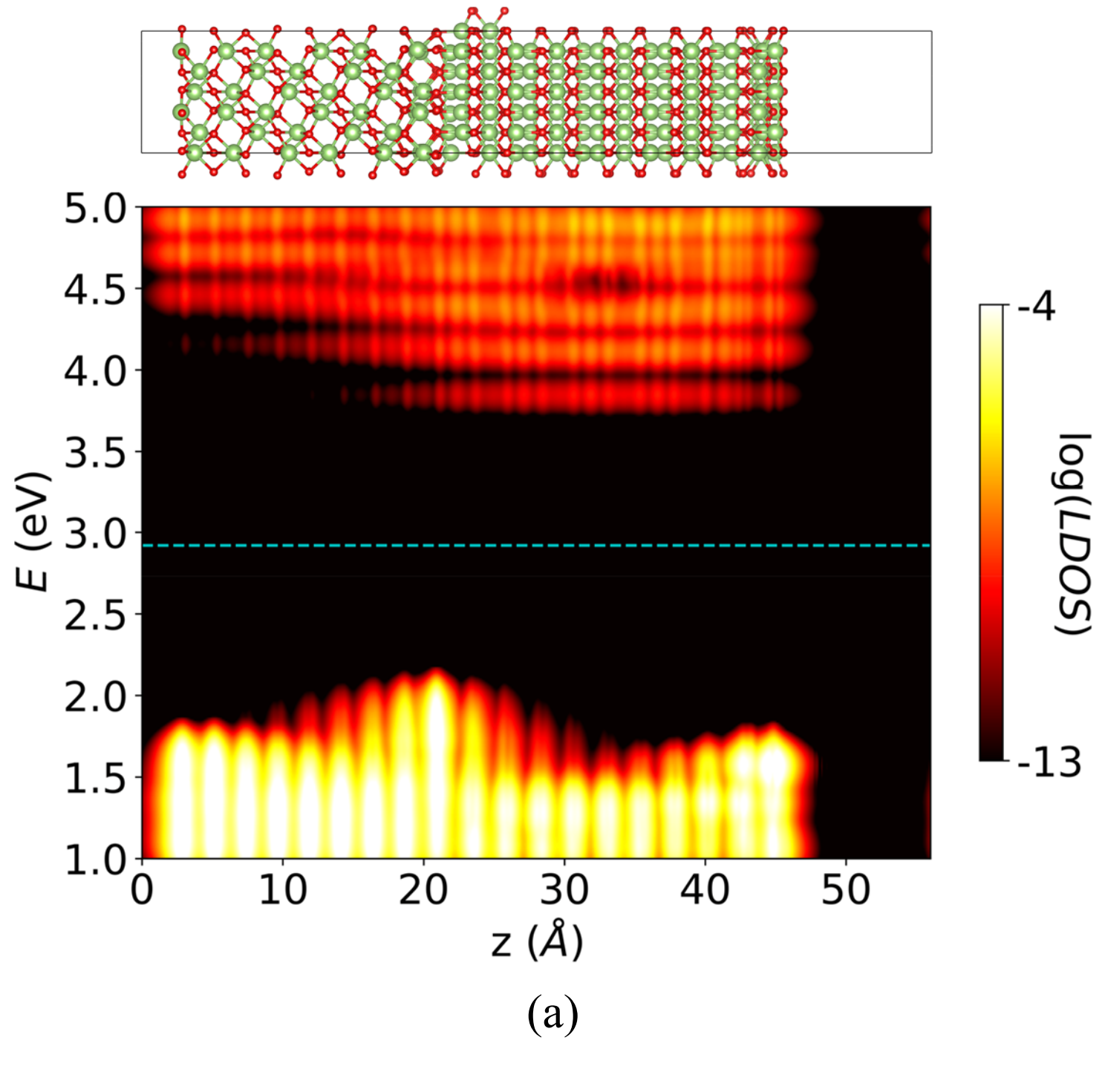
5.0
4.5
4.0
3.5
3.0
2.5
2.0
1.5
1.0
E (eV)
0
10
20
30
40
50
z (Å)
-4
-13
log(LDOS)

(a)

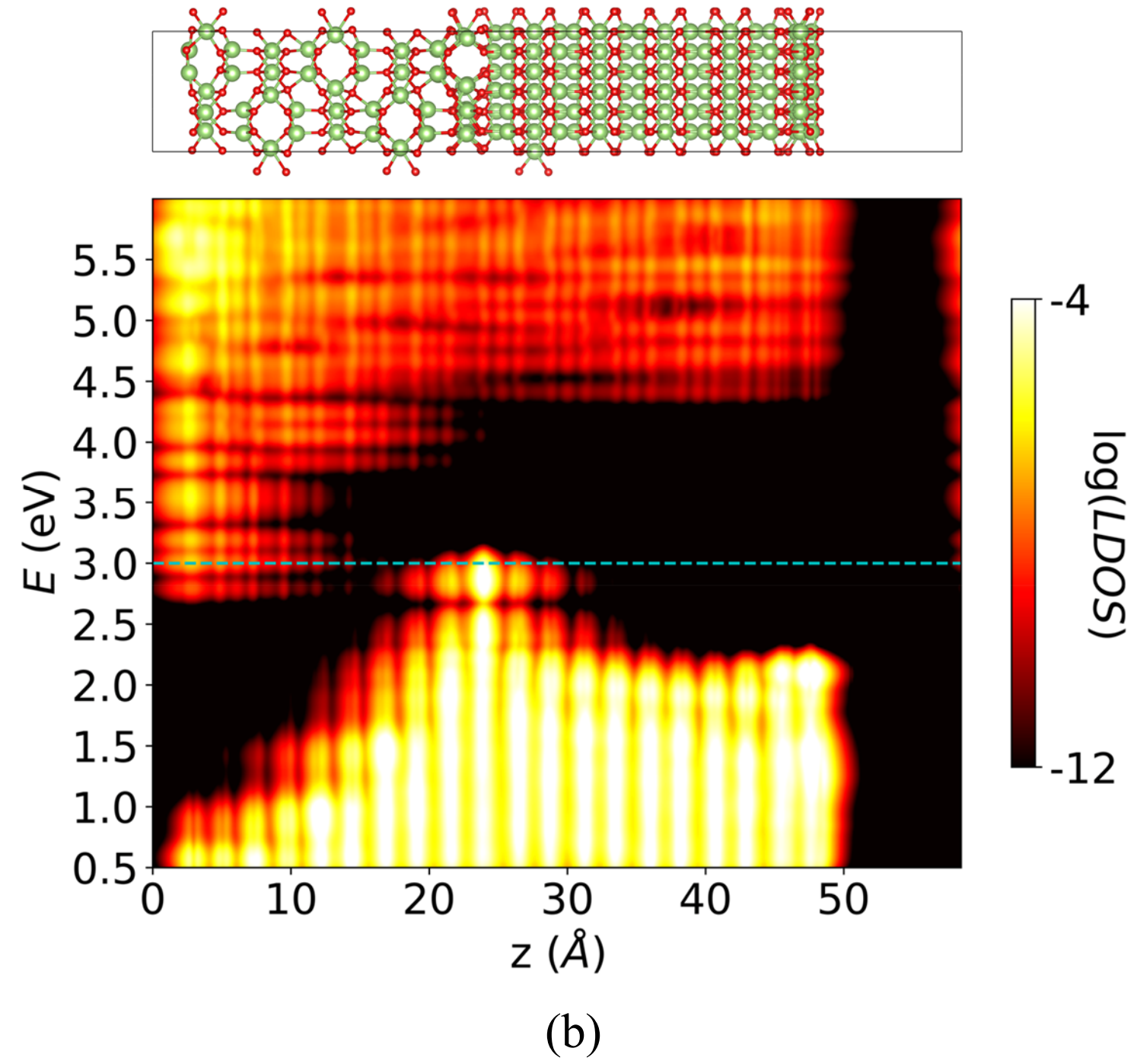
5.5
5.0
4.5
4.0
3.5
3.0
2.5
2.0
1.5
1.0
0.5
E (eV)
0
10
20
30
40
50
z (Å)
-4
-12
log(LDOS)

(b)

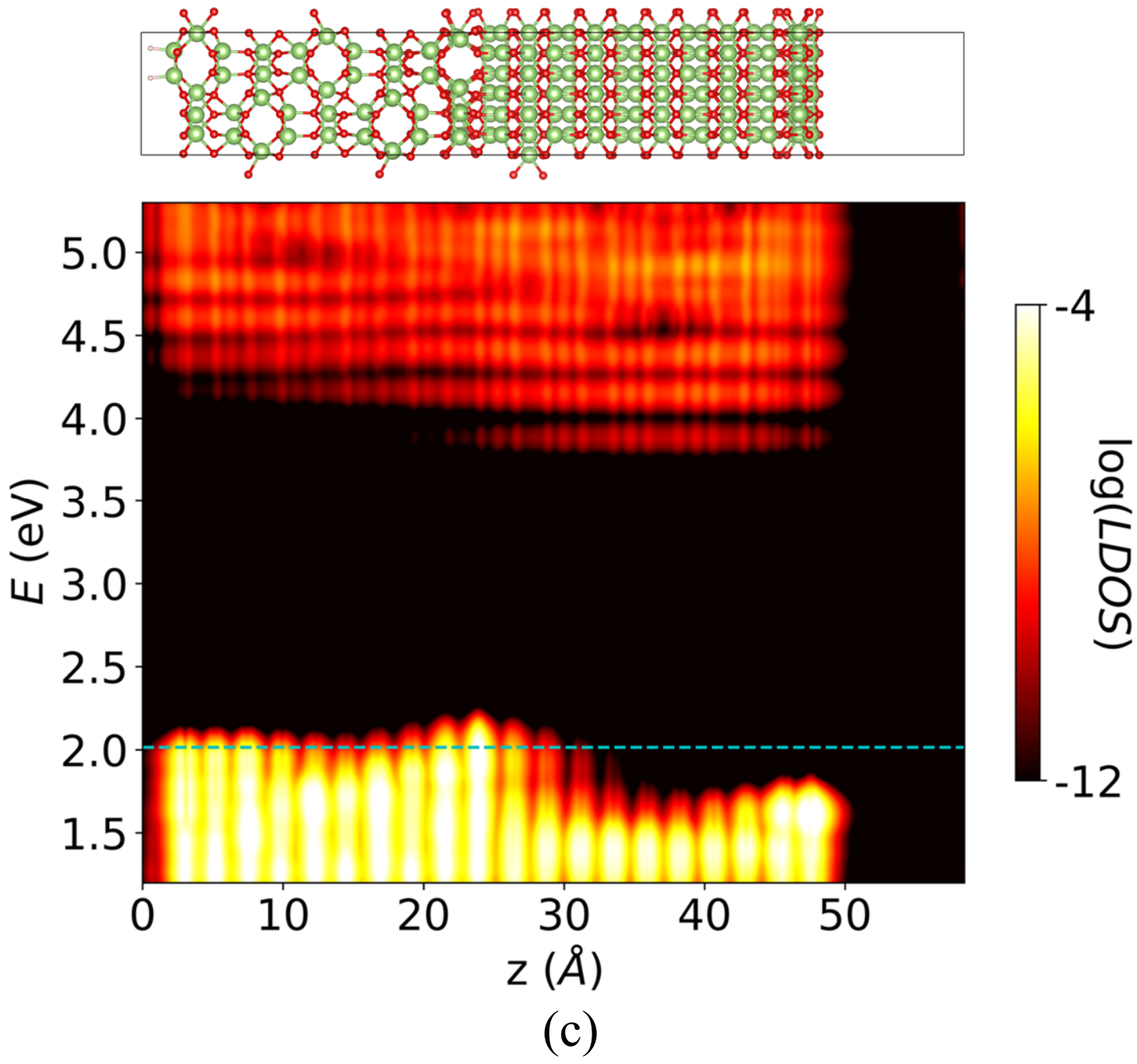

(c)

Figure 5. In-plane integrated LDOS calculated using the PBEsol functional, a **k**-point mesh of 3×5×1, a smearing parameter of 0.05 eV, and an energy resolution of 0.01 eV for the α-$Ga_2O_3$/β-$Ga_2O_3$ (a), κ-$Ga_2O_3$/β-$Ga_2O_3$ (b), and pseudo-hydrogen passivated κ-$Ga_2O_3$/β-$Ga_2O_3$ (c) heterojunctions. The atomic structure of each slab is added on the top of each LDOS map. Cyan dashed lines indicate the energy positions of the Fermi levels.